%% file: ZgammaQED.tex
\documentclass[a4paper,12pt]{article}
\pdfoutput=1
\usepackage[english]{babel}
\usepackage{jheppub_mod}
\usepackage{booktabs} % Per linee orizzontali migliori
\usepackage{scalerel}
\usepackage{float}
\usepackage{xcolor}
\usepackage[T1]{fontenc}
\usepackage{mathrsfs}
\usepackage{lmodern}

\usepackage{amsmath,amssymb,graphicx,textcomp,enumerate,xspace}
\usepackage{soul} %Use \st from soul instead of \sout from ulem; ulem breaks the biblio
\usepackage[utf8]{inputenc}
\input{texmacros}

\usepackage{ulem}
\title{Towards the inclusion of NLO EW corrections in the MiNLO method in
  Drell-Yan processes}

\author[a]{Filippo Belloni,}
\author[a]{Mauro Chiesa,}
\author[b,c]{Carlo Oleari,}
\author[b,c,1]{Emanuele Re\note{On leave of absence from LAPTh, Universit\'e Grenoble Alpes,
 Universit\'e Savoie Mont Blanc, CNRS, F-74940 Annecy, France.}}

\emailAdd{filippo.belloni@pv.infn.it}
\emailAdd{mauro.chiesa@pv.infn.it}
\emailAdd{carlo.oleari@mib.infn.it}
\emailAdd{emanuele.re@mib.infn.it}

\affiliation[a] {INFN, Sezione di Pavia, via Bassi 6, 27100 Pavia, Italy}
\affiliation[b] {Universit\`a degli Studi di Milano\,-\,Bicocca,
  Piazza della Scienza 3, 20126 Milano, Italy}
\affiliation[c] {INFN, Sezione di Milano\,-\,Bicocca,
  Piazza della Scienza 3, 20126 Milano, Italy}
\preprint{LAPTH-010/26}

\abstract{
  In this paper we present the first application of the MiNLO method to the
  calculation of QED NLO corrections to the production of a neutral vector
  boson in Drell-Yan processes.  We consider only the case of initial-state
  radiation, when the $Z$ boson decays into neutrinos.
  We illustrate the abelianization procedure of the MiNLO formulae and
  discuss the impact that it has on the differential cross section.  We then
  propose a variant of the MiNLO formulae in order to circumvent some of the
  problems that arise when dealing with QED emissions.
  Since this is a case study, we use ad-hoc parton distribution functions and
  a larger value of the electromagnetic coupling constant, in order to
  emphasize potential discrepancies with respect to the expected behavior of
  the MiNLO formulae.
  We quantify the uncertainties connected with the proposed method, also for
  a the physical value of the electromagnetic coupling.
  The study presented here is a necessary first step towards incorporating
  full electroweak effects into the MiNNLO$_{\text{PS}}$ framework.  }


\begin{document}

\maketitle

%%%%%%%%%%%%%%%%%%%%%%%%%%%%%%%%%%%%%%%%%%%%%%%%%%%%%%%%%%%%%%
%%%%%%%%%%%%%%%%%%%%%%%%%%%%%%%%%%%%%%%%%%%%%%%%%%%%%%%%%%%%%%
\section{Introduction}
\label{sec:intro}
%%%%%%%%%%%%%%%%%%%%%%%%%%%%%%%%%%%%%%%%%%%%%%%%%%%%%%%%%%%%%%
%%%%%%%%%%%%%%%%%%%%%%%%%%%%%%%%%%%%%%%%%%%%%%%%%%%%%%%%%%%%%%

The charged and neutral Drell–Yan~(DY) processes constitute one of the
cornerstones of precision physics at hadron colliders. Notably, owing to its
clean experimental signature and high production rate, neutral current~(NC)
Drell-Yan production provides stringent tests of the Standard Model~(SM) and
plays a crucial role in precision measurements, such as the extraction of
electroweak~(EW) parameters and the calibration of detector response. On a
similar note, the charged-current~(CC) production is crucial, among other
reasons, for extracting the mass of the $W$ boson~($\MW$), which is one of
the fundamental parameters of the SM, and whose precise knowledge is of
significant importance to test the theory at the quantum level.

The accumulated statistics at the LHC, together with the effort put forth by
the experimental collaborations to reduce systematic uncertainties, have
pushed the precision of these measurements to an unprecedented level~(see
e.g.~Refs.~\cite{CMS:2024ony, ATLAS:2015ihy, LHCb:2015jyu} for the weak
mixing angle $\sin^2 \theta^{\rm \sss eff}_{\sss W}$, or~\cite{ATLAS:2024erm,
  CMS:2024lrd} for the $W$-boson mass), making theoretical uncertainties a
non-negligible component of the total error budget.  This is particularly
relevant in view of the High-Luminosity LHC~(HL-LHC) era, where, for
instance, the target precision for $\sin^2 \theta^{\rm \sss eff}_{\sss W}$ is
$15 \times 10^{-5}$, requiring, in turn, the theoretical simulation of the
forward-backward asymmetry to be under control at the level of few $10^{-4}$
in the peak region.  In this context, the need for increasingly accurate
theoretical predictions and reliable Monte Carlo simulation tools has become
particularly pressing. For instance, in measurements based on the template
fit method, theoretical predictions are used to generate templates as
functions of the parameters to be extracted, and any approximation in the
modeling~(from the truncation of the perturbative series to parametric
uncertainties) directly propagates into systematic uncertainties of
theoretical origin. This is especially relevant for electroweak precision
measurements based on NC and CC DY processes.

Although electroweak corrections are typically subleading at hadron
colliders, as a consequence of the hierarchy between the couplings $\alpha$
and $\alpha_s$ with $\alpha \sim \alpha_s^2$, it is well known that they
induce phenomenologically significant effects.  Photon radiation,
particularly from the final state, can lead to sizable distortions of key
kinematic distributions and induce radiative return effects in the presence
of resonances. Furthermore, at high invariant masses, EW Sudakov logarithms
and photon-induced subprocesses significantly enhance the impact of
electroweak corrections~(see Ref.~\cite{Denner:2019vbn} for a comprehensive
overview of the main features of electroweak corrections).  For the Drell-Yan
process, NLO EW corrections have been available for a long
time~\cite{Dittmaier:2001ay, Baur:2004ig, Zykunov:2006yb, Arbuzov:2005dd,
  CarloniCalame:2006zq, Baur:2001ze, Zykunov:2005tc, CarloniCalame:2007cd,
  Arbuzov:2007db, Dittmaier:2009cr, Baur:1998kt}, and the effects of multiple
photon radiation have been extensively studied~\cite{Placzek:2003zg,
  CarloniCalame:2003ux, CarloniCalame:2007cd, Brensing:2007qm,
  Dittmaier:2009cr}. These NLO EW corrections are currently available in both
dedicated Monte Carlo codes~\cite{CarloniCalame:2006zq, CarloniCalame:2007cd,
  Baur:1997wa, Baur:2001ze, Arbuzov:2007db, Andonov:2008ga, Bardin:2012jk,
  Bondarenko:2013nu, Arbuzov:2015yja, Arbuzov:2016wfy, Arbuzov:2022sep,
  Dittmaier:2009cr, Campbell:2016dks} and automated
tools~\cite{Alwall:2014hca, Frederix:2018nkq, Biedermann:2017yoi,
  Sherpa:2024mfk}.  Tuned comparisons of the theory predictions for CC and NC
DY processes obtained with some of these tools can be found in
Ref.~\cite{Alioli:2016fum}.  In most of these programs, the weak corrections
are combined additively with NLO QCD results. More recently,
next-to-next-to-leading order~(NNLO) QCD corrections have been combined with
NLO EW effects in codes such as FEWZ~\cite{Li:2012wna} and
MATRIX~\cite{Grazzini:2017mhc, Buonocore:2019puv, Grazzini:2019jkl}.
Furthermore, several tools incorporate the contribution of the leading
fermionic corrections beyond NLO, while the full two-loop EW corrections,
expressed in terms of form factors, have been made available in the GRIFFIN
library~\cite{Chen:2022dow}.  The exclusive exponentiation of QED correction
for DY processes within the YFS framework is implemented in the event
generator {\tt KKMC-hh}~\cite{Jadach:2016zsp, Jadach:2017sqv,
  Yost:2019bmz,Yost:2020jin, Yost:2022kxg}.  Significant progress has also
been achieved in the computation of mixed QCD-EW corrections, which have been
shown to reach the percent level for observables relevant to present and
future LHC analyses, competing with residual perturbative uncertainties from
higher-order QCD effects~\cite{Delto:2019ewv, Cieri:2020ikq,
  Bonciani:2016wya, Bonciani:2019nuy, Bonciani:2020tvf, Bonciani:2021iis,
  Buccioni:2020cfi, Behring:2020cqi, Dittmaier:2014qza, Dittmaier:2015rxo,
  Dittmaier:2020vra, Dittmaier:2024row, deFlorian:2018wcj, Bonciani:2021zzf,
  Buccioni:2022kgy, Armadillo:2022bgm, Buonocore:2021rxx, Armadillo:2024nwk,
  Behring:2021adr}.  Progress on the two-loop electroweak corrections to NC
DY is documented in~\cite{Armadillo:2025mfx}.  In addition to fixed-order
calculations, significant progress has recently been achieved in
incorporating QED effects into the resummation of large logarithmic
corrections to the vector-boson transverse momentum spectrum. Mixed
QCD$\otimes$QED resummation at next-to-leading logarithmic accuracy~(NLL) +
NLO was developed in Refs.~\cite{Cieri:2018sfk, Autieri:2023xme} for on-shell
$W$ and $Z$ production, and in Ref.~\cite{Buonocore:2024xmy} for off-shell
dilepton production with massive fermions. The resummation of mixed
QCD$\otimes$QED at next-to-next-to-leading-logarithmic~(NNLL) accuracy for NC
DY was performed in~\cite{Autieri:2025tho}.\footnote{Related work on mixed
QCD$\otimes$QED Sudakov resummation can also be found in
Ref.~\cite{Billis:2019evv}.}

The steadily increasing experimental accuracy requires theoretical
predictions that consistently incorporate higher-order radiative corrections
beyond QCD and, ideally, combine fixed-order~(FO) calculations with parton
shower~(PS) simulations. While NLO QCD corrections and their matching to
parton showers~(NLO+PS) are by now standard and essentially
automated~\cite{Nason:2004rx, Frixione:2007vw, Alioli:2010xd, Jezo:2015aia,
  Frixione_2002, Alwall:2014hca, Bellm:2015jjp, Platzer:2011bc,
  Hoeche:2011fd, Sherpa:2019gpd, Bauer:2008qh, Alioli:2013hqa}, the matching
of FO electroweak effects with PS has received comparatively less attention.
%, despite NLO EW corrections to DY productin have been known for a
%relatively long time, as outlined above. To date, only a limited
While several studies have explored the approximate inclusion of EW
corrections within higher-order QCD frameworks~\cite{Kallweit:2015dum,
  Gutschow:2018tuk, Brauer:2020kfv, Lindert:2022qdd}, the consistent matching
of NLO electroweak corrections to parton showers remains, to date, limited to
only a few specific results~\cite{Barze:2012tt, Bernaciak:2012hj,
  Barze:2013fru, Muck:2016pko, Chiesa:2019nqb, Granata:2017iod,
  Chiesa:2020ttl, Jager:2022acp, Chiesa:2019ulk}.  Such studies have so far
been primarily focused on Drell-Yan production processes, and, often, the
matching is obtained simultaneously for QCD and EW corrections in the POWHEG
framework.  By consequence, such Monte Carlo generators also include, within
some approximation, the factorizable subset of the mixed QCD-EW corrections
to inclusive DY production.  In particular, the code in
Ref.~\cite{Barze:2012tt} was used to study the theory uncertainties from weak
and mixed QCD-EW effects on the determination of the $W$-boson mass at the
LHC~\cite{CarloniCalame:2016ouw}.

At the same time, DY-based measurements are becoming increasingly
differential with respect to kinematic variables such as the transverse
momentum of the vector boson~($\pt$), which is known to be critically
sensitive to QCD initial-state radiation~(ISR) and plays a crucial role, for
instance, in $W$-mass measurements. In current simulation strategies, EW
effects are often included using tools that provide only leading-order
accuracy in the description of the vector-boson transverse momentum, while
the latter is modeled using higher-order QCD calculations that do not include
EW corrections. This mismatch highlights the importance of developing
simulation tools capable of simultaneously providing an accurate description
of both inclusive observables and radiation-sensitive quantities,
consistently combining state-of-the-art QCD and EW predictions.  The
development of event generators at this level of accuracy would require,
among other ingredients, the inclusion of NLO EW corrections~(and
specifically NLO QED) not only for the inclusive Drell-Yan process, but also
for the radiative process DY$+\ph$, where $\ph$ denotes a resolved emission
at $\mathcal{O}(\alpha)$, which can consist either of a photon or an
(anti-)quark originating from an $\mathcal{O}(\alpha)$ splitting. The latter
contribution is necessary to describe the $\pt$ spectrum with NLO accuracy in
the high transverse-momentum region. A further essential ingredient is the
merging of NLO predictions for the inclusive process with those for the
process featuring one resolved radiation.  Despite the advanced technology
developed for the DY process and the fact that electroweak corrections for
$W/Z+\gamma$ and $W/Z+j$ are known\footnote{The issue of matching QCD
corrections with parton showers in the presence of isolated photons has been
addressed at NLO for the process $pp \to W(l\nu)\gamma$ and direct $\gamma$
production in Refs.~\cite{Barze:2014zba, Jezo:2016ypn} and at NNLO for
$Z+\gamma$, $W+\gamma$, and $\gamma\gamma$ production in
Refs.~\cite{Lombardi:2020wju, Cridge:2021hfr,
  Gavardi:2022ixt}.}~\cite{Denner:2014bna, Denner:2015fca}, a unified
framework capable of consistently describing both unresolved and resolved QED
emission has been lacking so far.

In this work, we take a first step towards addressing this issue by
constructing an event generator that merges DY production including NLO QED
corrections with DY plus one resolved QED radiation at NLO QED accuracy,
employing the MiNLO method\footnote{In Ref.~\cite{Hamilton:2012rf} the
improved MiNLO method proposed originally in~\cite{Hamilton:2012np} was
dubbed MiNLO$^\prime$, in order to distinguish the two formulations.  In the
MiNLO$^\prime$ framework, one also recovers NLO accuracy, for observables
inclusive with respect to the emitted radiation.  In this paper, for ease of
notation, we generically adopt the notation MiNLO instead of
MiNLO$^\prime$.}~\cite{Hamilton:2012np, Hamilton:2012rf}. The resulting
computation achieves NLO accuracy in the resolved radiation regime, while
simultaneously reproducing NLO predictions for inclusive DY observables.  The
MiNLO method was originally formulated to obtain predictions for the process
$pp \to F+j$~($F$ being a colorless system) at NLO QCD accuracy, with the
additional feature of recovering the NLO accurate results for the inclusive
$pp\to F$ process, when observables inclusive with respect to radiation are
considered. The algorithm served also as the starting point for the
formulation of the MiNNLO method~\cite{Monni:2019whf, Monni:2020nks}, which
allows simulating the $F+j$ process at NLO accuracy in QCD, while achieving
NNLO accuracy for the inclusive $pp\to F$ process~(NNLO+PS). Notably, the
MiNNLO approach, and, de facto, MiNLO, was also generalized to the case of
heavy colored final states, such as for $t\bar{t}$ production in hadronic
collisions~\cite{Mazzitelli:2020jio}, as well as to other final states
featuring two heavy quarks~\cite{Mazzitelli:2023znt, Mazzitelli:2024ura,
  Biello:2024pgo}.  Other algorithms for the matching of NNLO QCD
calculations to parton showers have been developed in~\cite{Hamilton:2013fea,
  Alioli:2013hqa, Hoche:2014uhw, Hoche:2014dla}.

The adaptation of the MiNLO method to QED represents an essential step
towards a unified treatment of QCD and electroweak corrections within modern
event generators and lays the groundwork for future extensions aiming at NLO
QCD + NLO EW accuracy~(with matching to QCD and QED parton shower) for finite
values of the vector boson $\pt$, and ultimately NNLO QCD + NLO EW
precision~(plus parton shower) for observables inclusive with respect to
strong and electromagnetic radiation.

In the following sections, we will attempt to follow the original MiNLO
formulation for QCD as closely as possible to highlight the new aspects and
criticalities of its application to QED, using the $pp \to Z \ph \to
\nu\bar{\nu}\ph$ process~(where $\ph=\{q,\bar{q},\gamma\}$ and the
initial-state particles can be quarks and photons) as a case study, in order
to concentrate on a process with only initial-state radiation.\footnote{This
is the first step towards the more general case of massive charged-lepton
production $pp \to Z \ph \to \mu^+\mu^-\ph$, where also final-state radiation
is present, following the work done for $t\bar{t}$ production of
Ref.~\cite{Mazzitelli:2020jio}.}  At this stage, we will not make any attempt
to interface our MiNLO process with a parton shower, since the result would
be unphysical without the inclusion of the QCD radiation.  In addition,
current parton-shower tools for LHC physics are not designed to treat only
QED emissions, without QCD effects.

The paper is organized as follows. In Sec.~\ref{sec:minlo} we recall the
formulation of the MiNLO method for QCD.  The key aspects and the main issues
of the adaptation of the MiNLO algorithm to the case of QED corrections are
discussed in Sec.~\ref{sec:minlo_QED}, where we describe the treatment of the
QED coupling constant and detail the abelianization of the MiNLO formulas as
well as the treatment of the parton distribution functions.  The position of
the Sudakov peak is also discussed together with its consequences on the
formulation of the method. We present the technical details of the
implementation of our calculation in Sec.~\ref{sec:implementation}. The
validation of the calculation and a discussion of the numerical results can
be found in Sec.~\ref{sec:res}. Finally we give our conclusions in
Sec.~\ref{sec:conclusions}.

\section{The MiNLO formalism for QCD}
\label{sec:minlo}

In this section, we recall the main formulae underlying the MiNLO method.  We
use the the notation of Ref.~\cite{Monni:2019whf} for the QCD case, applied
here to the $pp\to Z(\to\nu\bar{\nu})\,X$ process at NLO accuracy in QCD, where
$X$ denotes the radiation against which the $Z$ recoils.
We will then extend the formalism to the case of QED corrections,
illustrating all the issues related with the abelianization procedures.

Up to $\mathcal{O}(\as^2)$, the differential cross section as a function of
the Born phase space for $Z$ production, $\PhiB$, and of the transverse
momentum of the $Z$ boson, $\pt$, can be written as\footnote{For ease of
notation, we do not indicate explicitly the $\PhiB$ dependence of ${\cal L}$
and $R_f$.}
\begin{equation}
  \label{eq:ptres}  
  \frac{\mathd\sigma}{\mathd\PhiB\mathd \pt}  = \frac{\mathd\sigma^{\rm
      sing}}{\mathd\PhiB\mathd \pt} +R_f\!\(\pt\),
\end{equation}
where $R_f$ contains terms that are non-singular in the small $\pt$ limit
and the singular differential cross section can be written as
\begin{equation}
  \label{eq:sigma_sing}
\frac{\mathd\sigma^{\rm sing}}{\mathd\PhiB\mathd \pt} =  \frac{\mathd}{\mathd \pt}
  \lg \exp\!\lq -\tilde{S}\!\(\pt\) \rq \! \Lum\!\( \pt\)\rg ,
\end{equation}
where $\tilde{S}\!\(\pt\) $ is the Sudakov exponent and $\Lum\!\( \pt\)$
involves the parton luminosities, the Born squared amplitude for the
production of the $Z$ boson, the hard-virtual corrections and the collinear
coefficient functions (see Ref.~\cite{Monni:2019whf} for more details).
We can further elaborate eq.~(\ref{eq:sigma_sing}) and write
eq.~(\ref{eq:ptres}) as
\begin{equation}
\label{eq:ptres1}  
\frac{\mathd\sigma}{\mathd\PhiB\mathd \pt}  = 
 \exp\!\lq -\tilde{S}\!\(\pt\) \rq D(\pt)+R_f\!\(\pt\),
 % \equiv \frac{\mathd\sigma^{\rm sing}}{\mathd\PhiB\mathd \pt}+R_f\!\(\pt\),
\end{equation}
where 
\begin{equation}
  \label{eq:D}
  D(\pt) \equiv -\frac{\mathd \tilde{S}\!\(\pt\)}{\mathd \pt}\Lum\!\(\pt\)
  +\frac{\mathd \Lum\!\(\pt\)}{\mathd \pt}.
\end{equation}
\dontshow{
Up to $\mathcal{O}(\as^2)$, the differential cross section as a function of
the Born phase space for $Z$ production, $\PhiB$, and of the transverse
momentum of the $Z$ boson, $\pt$, can be written as\footnote{For ease of
notation, we do not indicate explicitly the $\PhiB$ dependence of ${\cal L}$
and $R_f$.}
\begin{equation}
  \label{eq:ptres}  
  \frac{\mathd\sigma}{\mathd\PhiB\mathd \pt}  =  \frac{\mathd}{\mathd \pt}
  \lg \exp\!\lq -\tilde{S}\!\(\pt\) \rq \! \Lum\!\( \pt\)\rg +R_f\!\(\pt\),
\end{equation}
where $R_f$ contains terms that are non-singular in the small $\pt$ limit. We
can further elaborate this formula and write it as
\begin{equation}
\label{eq:ptres1}  
\frac{\mathd\sigma}{\mathd\PhiB\mathd \pt}  = 
 \exp\!\lq -\tilde{S}\!\(\pt\) \rq D(\pt)+R_f\!\(\pt\)
  \equiv \frac{\mathd\sigma^{\rm sing}}{\mathd\PhiB\mathd \pt}+R_f\!\(\pt\),
\end{equation}
where 
\begin{equation}
  \label{eq:D}
  D(\pt) \equiv -\frac{\mathd \tilde{S}\!\(\pt\)}{\mathd \pt}\Lum\!\(\pt\)
  +\frac{\mathd \Lum\!\(\pt\)}{\mathd \pt}\,.
\end{equation}
}
The Sudakov exponent and its derivative are given by
\begin{equation}
  \label{eq:Sud}
  \tilde{S}\!\(\pt\) = 2\int_{\pt}^{Q}\frac{\mathd q}{q}
  \lg  A\!\( \as\!\(q\)\) \log\frac{Q^2}{q^2} +
  \tilde{B}\!\( \as\!\(q\)\) \rg,
\end{equation}
and
\begin{equation}
  \label{eq:DSud}  
  \frac{\mathd \tilde{S}\!\(\pt\)}{\mathd \pt} =
  - \frac{2}{\pt } \lg
   A\!\( \as\! \(\pt\)\) \log\frac{Q^2}{\pt^2} +
  \tilde{B}\!\(  \as\!\(\pt\)\) \rg ,
\end{equation}
where $Q$ is the virtuality of the $Z$ boson.
In the specific case of Drell-Yan production, the functions $A$ and $B$, up
to second order in $\as$, have the following expressions
\begin{eqnarray}
  \label{eq:AB}
  A(\as) &=& \left(\abar\right) A^{(1)} + \left(\abar\right)^2 A^{(2)}\,,
  \nonumber \\
  \tilde{B}(\as) &=& \left(\abar\right) B^{(1)} + \left(\abar\right)^2 \tilde{B}^{(2)}\,,
\end{eqnarray}
where 
\begin{eqnarray}
  \label{eq:ABcoeff}
  A^{(1)}&=&2 \CF, \qquad \qquad \qquad B^{(1)}=-3\CF, \nonumber \\
  A^{(2)}&=& \left( \frac{67}{9} - \frac{\pi^2}{3} \right)\CA \CF -
  \frac{20}{9}\CF \TF \nf,
  \nonumber \\ 
  B^{(2)}&=&\left(-\frac{17}{12} - \frac{11 \pi^2}{12} +  6 \zeta_3\right) \CA \CF \nonumber \\
  & & \mbox{} +  \left( -\frac{3}{4} + \pi^2 - 12\zeta_3 \right) \CF^2+
  \left( \frac{1}{3}+\frac{\pi^2}{3} \right) \CF \TF \nf \nonumber\\ 
  \tilde{B}^{(2)} &=& B^{(2)} + 2\pi\,\beta_0\, H + 2\zeta_3\(A^{(1)}\)^2, \qquad\qquad
  \beta_0 = \frac{11\CA-4\TF\nf}{12\pi}\,,
\end{eqnarray}
$H$ being the hard-virtual coefficient function
\begin{equation}
\label{eq:Hcoeff}
  H = \left( -8 + \frac{7}{6}\pi^2 \right) \CF\,.
\end{equation}
The non-singular term $R_f$ in eq.~(\ref{eq:ptres}) can be written as
\begin{equation}
R_f\!\(\pt\) = \frac{\mathd\sigma^{\rm \sss (NLO)}_{\scriptscriptstyle\rm ZJ}}{\mathd\PhiB\mathd
      \pt}-
  \abarmu{\pt}\left[\frac{\mathd\sigma^{\rm sing}}{\mathd\PhiB\mathd
      \pt}\right]^{(1)}
  - \(\abarmu{\pt}\right)^2\left[\frac{\mathd\sigma^{\rm sing}}{\mathd\PhiB\mathd
      \pt}\right]^{(2)}\!\!,
  \label{eq:Rf}
  \end{equation}
where 
\begin{equation}
\frac{\mathd\sigma^{\rm (\sss NLO)}_{\scriptscriptstyle\rm ZJ}}{\mathd\PhiB\mathd
  \pt} = \abarmu{\pt}\left[\frac{\mathd\sigma_{\scriptscriptstyle\rm
      ZJ}}{\mathd\PhiB\mathd \pt}\right]^{(1)} +
\left(\abarmu{\pt}\right)^2\left[\frac{\mathd\sigma_{\scriptscriptstyle\rm
      ZJ}}{\mathd\PhiB\mathd \pt}\right]^{(2)}\,
\end{equation}
is the differential cross-section at fixed NLO accuracy for $Z+1$ jet
production. In writing these expressions, we are using the notation
$[X]^{(i)}$ to denote the $i$-th component of the expansion of $X$ in powers
of $\as/2\pi$.

The implementation of the MiNLO formula in eq.~(\ref{eq:ptres1}) is then given by
\begin{eqnarray}
  \label{eq:STDminlo}  
  \frac{\mathd\sigma}{\mathd\PhiB\mathd \pt}  &=& 
  \exp\lq-\tilde{S}(\pt)\rq
  \lg \abarmu{\pt}\left[\frac{\mathd\sigma_{\scriptscriptstyle\rm ZJ}}{\mathd\PhiB\mathd
      \pt}\right]^{(1)} \left(1+\abarmu{\pt} [\tilde{S}(\pt)]^{(1)}\right) \right.
  \nonumber\\
 && \phantom{\exp\lq-\tilde{S}(\pt)\rq\, \Big\{} +
\left.  \left(\abarmu{\pt}\right)^2\left[\frac{\mathd\sigma_{\scriptscriptstyle\rm
        ZJ}}{\mathd\PhiB\mathd 
    \pt}\right]^{(2)}\rg
\nonumber \\
&\equiv&
 \frac{\mathd\sigma^{\rm \sss PWG}}{\mathd\PhiB\mathd \pt}\,,
\end{eqnarray}
and this term corresponds to the MiNLO $\tilde{B}$ function in the {\tt
  POWHEG} language~\cite{Frixione:2007vw}.
%We will refer to this formula as $\mathd\sigma^{\rm \sss PWG}$ in the rest of
%the paper.

In the following, we also need the expansion in~$\as$ of $D(\pt)$ and of
$\lum$ in eq.~(\ref{eq:D})
\begin{eqnarray}
 D(\pt) &=& \(\frac{\as}{2\pi}\)[D(\pt)]^{(1)} + \(\frac{\as}{2\pi}\)^2
 [D(\pt)]^{(2)} + \ldots
 %\nonumber\\
 \\
     \lq D(\pt)\rq^{(1)} &=&
     -\left[\frac{\mathd \tilde{S}(\pt)}{\mathd \pt}\right]^{(1)}[{\cal L}(\pt)]^{(0)}
     + \left[\frac{\mathd {\cal L}(\pt)}{\mathd \pt}\right]^{(1)}\!\!\!\!,
     %\nonumber\\
     \\
     %  [D(\pt)]^{(2)}
     \label{eq:D2}
 \lq D(\pt) \rq^{(2)}  &=& -\left[\frac{\mathd \tilde{S}(\pt)}{\mathd \pt}\right]^{(2)}[{\cal L}(\pt)]^{(0)}
  -\left[\frac{\mathd \tilde{S}(\pt)}{\mathd \pt}\right]^{(1)}[{\cal
    L}(\pt)]^{(1)}+ \left[\frac{\mathd {\cal L}(\pt)}{\mathd
      \pt}\right]^{(2)} \!\!\!\!,\phantom{aa}
  \end{eqnarray}
where 
\begin{eqnarray}
  \label{eq:ll0}
  [{\cal L}(\pt)]^{(0)}  &=& \sum_{ij}\frac{\mathd|M^{\scriptscriptstyle\rm
      Z}|_{ij}^2}{\mathd\PhiB}\,f_i^{[a]}f_j^{[b]}\,, 
\\
  \label{eq:ll1}
  [{\cal L}(\pt)]^{(1)} & = & \sum_{ij}\frac{\mathd|M^{\scriptscriptstyle\rm
      Z}|_{ij}^2}{\mathd\PhiB  }\bigg\{H \, f_i^{[a]}f_j^{[b]}
  +\(C_{iq}\otimes f_q\)^{[a]}f_j^{[b]}+(C_{ig}\otimes f_g)^{[a]}f_j^{[b]}
  \nonumber  \\
  && \phantom{\sum_{ij}\frac{\mathd|M^{\scriptscriptstyle\rm
      Z}|_{ij}^2}{\mathd\PhiB  }\bigg\{H \, f_i^{[a]}f_j^{[b]}\,}
  + f_i^{[a]} (C_{jq}\otimes f_q)^{[b]}+f_i^{[a]}  (C_{jg}\otimes
  f_g)^{[b]}\bigg\}.
  \phantom{aaa}
\end{eqnarray}
Here $M^{\scriptscriptstyle\rm Z}$ is the Born matrix element for the
Drell-Yan production of the $Z$ boson, $[a]$ and~$[b]$ label the incoming partons, $i$
and~$j$ stand for the quark flavors in the $q\bar{q} \to Z$ process,
$f^{[a/b]}_{(q/g)}$ are the PDFs, while $C_{k\ell}(z)$ are the hard-collinear
functions up to $\ord{\as}$
\begin{equation}
  \label{eq:Cfuncs}
C_{k\ell}(z)= - \hat P_{k\ell}^{\epsilon}(z) - \frac{\pi^2}{12}\, \CF \,\delta_{k\ell}\,\delta(1-z)\,,
\end{equation}
where $\hat P_{k\ell}^{\epsilon}(z)$ is the $\ord{\epsilon}$ part of the LO
regularized Altarelli-Parisi splitting functions $\hat P_{k\ell}(z)$
\begin{align}
  \label{eq:Pfuncs}
  \hat P^{(0)}_{qq}(z) & = \CF\left[\frac{1+z^2}{(1-z)_+}+\frac32\delta(1-z)\right],
&  \hat P^{\epsilon}_{qq}(z) &= -\CF (1-z)\,,
  \nonumber\\[2mm]
  \hat P^{(0)}_{qg}(z) &= \TF \! \left[z^2+(1-z)^2\right]\,,
&
  \hat P^{\epsilon}_{qg}(z) &= -2\,\TF z(1-z)\,,
\end{align}
and we have used the notation
\begin{equation}
(f\otimes g)(x) = \int_x^1 \frac{\mathd z}{z} f(z) \,
g\!\left(\frac{x}{z}\right). 
\end{equation}
The symbol $(\cdots)_{+}$ denotes the usual plus prescription.
Equations~(\ref{eq:ll0}) and~(\ref{eq:ll1}) depend explicitly of the
phase space of the LO Drell-Yan process, while the dependence on $\pt$ is
introduced via the PDF factorization scale choice $\muf=\pt$.

\section{The MiNLO formalism for QED}
\label{sec:minlo_QED}

%\subsection{The Abelianization procedure and the Sudakov peak in QED}
%\label{sec:sud}
In this work, we want to extend the MiNLO method applied up to now to deal
with NLO QCD corrections, to NLO QED, in the simplest process, i.e.~$pp \to Z
\ph$, where $\ph$ represents any radiation allowed in QED from the underlying
inclusive process ($pp \to Z\to \nu\bar{\nu}$), namely:
$\ph,p =\{q,\bar{q},\gamma\}$.  Specifically, this implies that we neither fix
the flavor of the particle produced in association with the $Z$ boson, nor do
we require an isolated photon.  Instead, we address $Z+$jet-type processes
using a QED analogue of a jet.  For the same reasons, the particles in
the initial state can be quarks or photons.

Since we are following, as closely as possible, the QCD MiNLO formulation for
$F+$jet production, with $F$ colorless, in order to apply it to the QED case,
we limit ourselves to considering $Z$ decay into neutrinos, to prevent
final-state radiation~(FSR) emissions from the decay products.  The inclusion
of FSR from leptons could be obtained by following the MiNLO formulation for
the $t\bar{t}$+jet process described in Refs.~\cite{Mazzitelli:2020jio,
  Mazzitelli:2021mmm}, provided that massive leptons are considered. This is
left to future developments.

Adhering closely to the original QCD MiNLO formalism also requires addressing
two additional aspects specific to the QED case: the running of the
electromagnetic coupling constant and the replacement of the color factors
with their QED counterparts. Both aspects play a crucial role in determining
the position of the Sudakov peak in QED, and we discuss them in the
following.

% Since we want to follow the original MiNLO formalism for QCD as closely as
% possible, we have to deal with two issues: the running of the electromagnetic
% coupling constant and the replacement of the color factors with the
% corresponding QED ones. These two last points have a deep impact in the
% postion of the Sudakov peak in QED. We discuss each of these points in the following.

\subsection{The running of the electromagnetic coupling}
\label{sec:alpha}
In the evaluation of the Feynamn diagrams, the coupling constant at the
vertexes of the $Z$-boson propagator with the initial-state hadronic current
and with the final-state neutrinos can be assumed as independent from the
coupling used by MiNLO when generating radiation.  For this reason, even if
these couplings can be written in terms of the electric charge, and
ultimately in terms of the electromagnetic coupling, they can be fixed at a
constant value of $\alpha$, independently of the value of $\alpha$ used for
the radiation, typically running.

As far as the generation of QED radiation is concerned (i.e.~the
$q\bar{q}\gamma$ vertexes), in the MiNLO formulae we have to
substitute $\as(\mu)$ with the electromagnetic running coupling
$\alpha(\mu)$ defined in the \MSB{} scheme. Indeed, the original MiNLO
prescription requires that the ISR splitting coupling that appears in
the tree-level process $pp\to Z\ph$ is evaluated in the $\MSB$ scheme
at the scale $\pt$. This same choice is consistently adopted for both
the NLO real and virtual corrections.
% As far as the generation of QED radiation is concerned (i.e.~the
% $q\bar{q}\gamma$ vertexes), in the MiNLO formulae we have to substitute
% $\as(\mu)$ with $\alpha(\mu)$, the \MSB{} electromagnetic running constant.
% We will give more details about the treatment of the powers of $\alpha$
% entering the process in Sec.~\ref{sec:alpha}.
% Following the original MiNLO method as closely as possible, we must evaluate
% the ISR splitting coupling that appears in the tree-level process $pp\to
% Z\ph$ in the $\MSB$ scheme and at the scale $\pt$. This same coupling is also
% used for the NLO real and virtual corrections.
%
After renormalization, the $q\bar{q}\gamma$ vertex develops a
contribution of the form 
\begin{equation}
\label{eq:alpha_ren}
\alpha(\mu)\(1+2\delta Z_e(\mu)+\delta Z_\gamma\)\,
\end{equation}
where $\delta Z_e(\mu)$ is the electric charge counterterm in the $\MSB$
scheme at the renormalization scale $\mu$, while $\delta Z_\gamma$ is the
$\mathcal{O}(\alpha)$ component of the Lehmann–Symanzik–Zimmermann factor for
the external photon.

In this paper we are interested in keeping an arbitrary scale $\mu$ in the
running of the electromagnetic coupling constant $\alpha$, without resorting
to $\alpha_{\sss 0}$, the Thompson value of the electromagnetic coupling.
The contribution in eq.~(\ref{eq:alpha_ren}) then develops large corrections
associated with the fermionic loops~(see e.g.~\cite{Kallweit:2017khh}). In
fact, if the fermions are treated as massive, then large logarithms of the
fermion masses appear.  If they are treated as massless, then the expression
is affected by the presence of infrared poles.  The correct treatment of
these poles implies that we have also to include, in the real corrections,
diagrams of the type $q\bar{q} \to Z q' \bar{q}'$ plus all their
crossings.\footnote{We have implemented the corresponding
  FKS~\cite{Frixione:1995ms, Frixione:1997np} subtraction terms and their
  integrated counterparts in the {\tt POWHEG BOX RES}, in order to deal with
  these QED singularities.}

In this paper we are considering a simplified case study, where only $u$ and
$d$ quarks are present, and the PDFs evolve only via QED.  The running of
$\alpha$ is then performed at two loops in QED, with the sole contributions
of up and down quarks.  The running is frozen below an arbitrary cutoff (of
the order of the electron mass), where $\alpha$ is set equal to
$\bar{\alpha}_{\sss 0}$.

Unless stated otherwise, in the following we will use $\bar\alpha_{\sss
  0}=0.04$ as the starting value for the running of $\alpha$, corresponding
to approximately five times its physical value. The rationale for this choice
is that, in investigating the application of MiNLO to QED, we aim to analyze
effects proportional to the electromagnetic coupling. An enhanced coupling
magnifies these QED effects, allowing us to obtain statistically significant
corrections with a moderate computational cost in the Monte Carlo
simulations. Nonetheless, the chosen value of $\bar\alpha_{\sss 0}$ remains
sufficiently small to keep the running of $\alpha$ moderate over the entire
range of scales probed by our setup, ensuring that the qualitative behavior
of the QED corrections is not drastically altered.

In a more realistic context where QED corrections are calculated alongside
QCD corrections, one could use the Standard Model running of $\alpha$
(following, for instance, Ref.~\cite{Erler:1998sy}), taking $\alpha(\MZ)$ in
the \MSB{} scheme as input and then evolving it to low scales down to the
minimum value of $\pt$ used for the calculation (of the order of a GeV),
thereby reducing the dependence of the running on the light degrees of
freedom (as done in Refs.~\cite{Amoroso:2023uux, Chiesa:2024qzd}).  The
possibility of reformulating MiNLO without using the \MSB{} scheme for the
QED coupling is left for future exploration.

\subsection{The color factors}
  
The coefficients $A^{(i)}$ and $B^{(i)}$ in eq.~(\ref{eq:ABcoeff}),
as well as the hard-virtual function $H$ in eq.~(\ref{eq:Hcoeff}), the
collinear coefficient functions in eq.~(\ref{eq:Cfuncs}) and the
corresponding splitting functions in eq.~(\ref{eq:Pfuncs}), must be
abelianized, with the following replacements
%  ,following, for example, Ref.~\cite{Autieri:2023xme}:
\begin{equation}
  \CA \to 0,  \qquad\CF \to Q_f^2, \qquad \TF \to \NC^f\, Q_f^2, \qquad \beta_0
  \to \beta_0^{\rm \sss QED}\,,
  \label{eq:conv}
\end{equation}
where $Q_f$ is the fermionic electric charge.
Here, $\beta_0^{\rm \sss QED}$ is related to the running of $\alpha(\mu)$ in
the $\MSB$: since we will limit ourselves for simplicity to a simplified
parton model involving only up and down quarks, we have
\begin{equation}
\beta_0^{\rm \sss QED}=-\frac{\(Q_u^2+Q_d^2\)}{\pi}\,.
\end{equation}

\subsection{The QED Sudakov peak and the implications for the MiNLO formulation}
The most striking difference with respect to QCD turns out to be the position
of the Sudakov peak.  From the expression for the $Z$ boson
transverse-momentum distribution in eqs.~(\ref{eq:ptres})
and~(\ref{eq:sigma_sing}), by observing that the most singular contributions
in the limit $\pt\to 0$ are associated with the derivative of $\tilde{S}$
(which generates terms of the form $1/\pt \log (Q^2/\pt^2)$) as can be seen
in eq.~(\ref{eq:DSud}), it is possible to estimate the position of the peak
in the $\pt$ spectrum by finding the maximum of $\mathd
\big[e^{-\tilde{S}(p_{\scaleto{\rm T}{3pt}})}\big]/ \mathd\pt$.  In QCD, this
procedure leads to estimating the spectrum peak from a few GeV to tens of
GeV, depending on the colorless system produced.
In QED the peak position is several orders of magnitude lower and well
outside the practically accessible region. In fact, one can easily see that,
for $\alpha$ fixed (and this is a very good approximation in QED), the
position of the peak is approximately given by
\begin{equation}
\pt^{\rm\sss peak }= Q \exp\( -\frac{\pi}{4\, Q_f^2\, \alpha}\).
\end{equation}
For example, for an up-type quark~($Q_f=2/3$), with $Q=\MZ$ and
$\alpha=\alpha_0\equiv 1/137$ we get $\pt^{\rm\sss peak }\approx 6.5 \times
10^{-104}$~GeV.
This fact mainly arises from the different behavior of the coupling
constant at small $\pt$: in QCD, $\as(\pt)$ is divergent when $\pt\to 0$,
while in QED $\alpha(\pt)$ tends to $\alpha_0$. In addition to this, the
numerical values of the QCD and QED couplings that appear in the exponential
have an important impact too, together with the differences among the values
of the coefficients $A^{(i)}$ and $B^{(i)}$ in the two theories.

\dontshow{
As described in
Sect.~\ref{sec:alpha}, most of the subsequent numerical studies will be
performed using an unphysical value of $\alpha$, corresponding to a starting
value for the $\MSB$ running of $\alpha$ at low scales approximately five
times larger than $1/137$.  Using such value of $\alpha$ modifies the
position of the peak but does not change the conclusions stated above.
}

The position of the Sudakov peak is crucial: in fact, as it stands,
eq.~(\ref{eq:ptres}) is such that its integral over $\pt$, between an
infrared cutoff $\Lambda$ (more precisely, the scale value where the Sudakov
form factor in eq.~(\ref{eq:sigma_sing}) vanishes) and $Q$, reproduces
the NLO inclusive cross section for $Z$ production.
While not conceptually problematic, its position is so low that any numerical
evaluation of the actual implementation of the MiNLO formula,
i.e.~eq.~(\ref{eq:STDminlo}), is unfeasible, since the Born, the real and the
virtual amplitudes entering this equation will be unstable and unreliable
when evaluated for such low values of the transverse momentum.\footnote{These
considerations are even more relevant when combining the QED and QCD
corrections, since, in this case, the choice of the infrared cutoff $\Lambda$
would be constrained by the condition that QCD remains in its perturbative
regime and $\Lambda$ should not go below $\ord{1}$~GeV.}

In order to circumvent this obstacle, we have adopted the  following strategy: the
$\pt$ spectrum in eq.~(\ref{eq:ptres}) is divided into two regions, below
and above a technical cutoff $\ptcut$,
\begin{equation}
  \frac{\mathd\sigma}{\mathd\PhiB\mathd \pt} =
  \frac{\mathd\sigma^{\sss <}}{\mathd\PhiB\mathd\pt} +
  \frac{\mathd\sigma^{\sss >}}{\mathd\PhiB\mathd \pt}\,,
\end{equation}
where, according to eqs.~(\ref{eq:sigma_sing})
and~(\ref{eq:STDminlo}),
\begin{align}
  \label{eq:sigma<}
  \frac{\mathd\sigma^{\sss <}}{\mathd\PhiB\mathd\pt} &\equiv  \frac{\mathd}{\mathd \pt}
 \lg \exp\!\lq -\tilde{S}\!\(\pt\) \rq \! \Lum\!\( \pt\)\rg +R_f\!\(\pt\)
 &{\rm for\ \ }   \pt < \ptcut \,,
 \\[2mm]
 \label{eq:sigma>}
 \frac{\mathd\sigma^{\sss >}}{\mathd\PhiB\mathd \pt} &\equiv
 \frac{\mathd\sigma^{\rm \sss PWG}}{\mathd\PhiB\mathd \pt} &{\rm for\ \ }
 \pt > \ptcut \,.
\end{align}
In the region above $\ptcut$, the original MiNLO method, appropriately
abelianized, is applied.

Recalling that the $R_f\!\(\pt\)$ term is non singular in the $\pt \to 0$
limit, we expect this term to behave as power corrections in $\pt$.  In the
region below $\ptcut$, we then assume that the regular terms $R_f\!\(\pt\)$
in eq.~(\ref{eq:sigma<}) can be neglected, and the differential cross section
can be expressed as an exact differential in $\pt$.  We can then integrate
over the $Z$ boson transverse momentum (that is nevertheless assumed as
small, since we are in the region $\pt<\ptcut$) and obtain
\begin{equation}
  \label{eq:integrated_sigma<}
  \frac{\mathd\sigma^{\sss <}}{\mathd\PhiB}= \exp\!\lq -\tilde{S}\!\( \ptcut \) \rq \!
  \lg \lq\Lum(\ptcut)\rq^{(0)}+\frac{\alpha(\ptcut)}{2\pi} \lq\Lum(\ptcut)\rq^{(1)} \rg,
\end{equation}
where we have expanded the luminosity $\Lum(\ptcut)$ in series of $\alpha$,
using the abelianization of eqs.~(\ref{eq:ll0}) and~(\ref{eq:ll1}).  In this
way, the kinematics of the events in the region $\pt<\ptcut$ is correctly
computed according to the expression in eq.~(\ref{eq:integrated_sigma<}), and
the NLO accuracy in the totally inclusive $Z$ boson production is reached.

The validity of having disregarded the non-singular term $R_f\!\(\pt\)$ in
  this region will be discussed in Sec.~\ref{sec:res}.

\subsection{The QED evolution of parton distribution functions}
\label{sec:pdfs}
Since we are considering the MiNLO QED corrections to $pp\to Z\ph$
production, we would need a PDF set evolved only with QED kernels.  This is
due to fact that there is an interplay between the factorization scale
$\muf$, used for the PDF calculation, and the terms in the MiNLO formulae,
that evolve the PDFs from the initial scale $\muf$, of the order of $\pt$, up
to the hard process scale, i.e.~the virtuality of the $Z$ boson.

If we were to use PDFs containing also QCD evolution effects without
including QCD corrections in the MiNLO formula, the residual scale dependence
would be large and would correspond to the one of a leading-order calculation
(in QCD).

Since in this paper we are only interested in presenting the QED MiNLO
formalism as a case study, we have built an ad-hoc PDF set, that we evolve
only with QED kernels, from very low scales up to the $Z$ mass.
We proceeded as follows: we used the latest version of the HOPPET
library~\cite{Salam:2008qg, Karlberg:2025hxk}, which allows to evolve a PDF
set, starting from a grid defined at a given reference scale, performing only
the QED evolution. To date, the QED evolution has been performed only at
leading order. The consequences of this will be further discussed in
Sec.~\ref{sec:res}.
In addition, we have modified the HOPPET library to eliminate the QCD
evolution and to lower the minimum $\muf$ value achievable in the evolution.
The running of $\alpha$ was also modified to match exactly what described in
Sec.~\ref{sec:alpha}.

The way we have obtained this ad-hoc PDF has a very large degree of
arbitrariness. Our goal was only to have a set with the quarks and photon
PDFs of comparable size, to put on the same footing diagrams initiated only
by quarks and those where a photon is present in the initial state.  In
particular, for the numerical analysis presented in this work, we started
with the {\tt NNPDF31\_nlo\_as\_0118\_luxqed} set~\cite{Manohar:2016nzj,
  Manohar:2017eqh, Bertone:2017bme}, at its minimum $q$ value~(1.65~GeV), and
we have used this set as input to the QED-only evolution supplied by HOPPET,
where we have set the starting scale for the evolution to be 0.01~GeV.

Since the PDFs obtained in this way exhibit a sort of transient at small
evolution scales (where the photon PDF grows very rapidly at the expense of
the quarks), we reiterated the above procedure by using this set to obtain a
new initial condition for the evolution (computed where we have an
equilibrium between the quarks and the photon content), and repeated the
evolution starting from $q=0.01$~GeV. The PDFs of the resulting set, although
they grow quite rapidly between 0.01 and $\ord{1}$~GeV, no longer show
important relative variations between the different quark flavors and the photon.

\section{Implementation of the QED MiNLO formulae}
\label{sec:implementation}

The calculation of the NLO QED corrections to the process $pp \to Z \ph$ has
been implemented within the {\tt POWHEG BOX RES}~\cite{Jezo:2015aia}
framework and utilizes external libraries such as
LHAPDF~\cite{Buckley:2014ana} and HOPPET~\cite{Salam:2008qg,
  Karlberg:2025hxk} for PDF management (see Sec.~\ref{sec:pdfs}) and
RECOLA~\cite{Actis:2012qn, Actis:2016mpe, Denner:2017vms, Denner:2017wsf,
  Denner:2016kdg} for the computation of tree-level and one-loop matrix
elements.

It is important to note that, since the strategy of our study is to follow
the original MiNLO implementation for QCD as closely as possible, quarks and
photons are treated exactly on equal footing in our calculation. This
specifically implies that the calculation includes the tree-level and virtual
matrix elements for all processes of the type $q\bar{q} \to Z \gamma$ and
$\gamma q (\bar{q}) \to Z q (\bar{q})$, while the real contributions consist
of the processes $q\bar{q} \to Z \gamma \gamma$, $\gamma q (\bar{q}) \to Z q
(\bar{q}) \gamma$, $\gamma \gamma \to Z q\bar{q}$ and $q\bar{q} \to Z q'
\bar{q}'$, plus crossings, with $q$ equal or different from $q'$ (see
also the discussion in Sec.~\ref{sec:alpha}). In our case study, we
limit ourselves to a simplified theory in which only up and down quarks
are present.

\dontshow{
The calculation relies on the Complex-Mass-Scheme for the treatment of the Z
resonance and employs a hybrid scheme ($\alpha_{\sss\MSB}(\mu)$, $\MW$, $\MZ$)
for the electroweak parameters. In reality, the masses and widths of the
electroweak bosons enter only the Z propagator and the $Zf\bar{f}$ vertices
without affecting the QED corrections. Furthermore, since the $Zf\bar{f}$
vertices do not contribute to the corrections considered, the $\alpha$ factor
associated with the $Z$-fermion coupling has been chosen as $\alpha^{\rm
  \sss OS}(\MZ)$ and is independent of the coupling of the $q\bar{q}\gamma$
splitting, which is instead evaluated in the $\MSB$ scheme according to the
MiNLO method (see Sec.~\ref{sec:alpha}).
}

\dontshow{
The calculation of the photonic corrections in the virtual diagrams has been
performed using RECOLA.  The fermionic loops enter only in the counterterm
of $\alpha(\mu)$ and in the $\delta Z_\gamma$ factor (in the notation of
Refs.~\cite{Denner:1991kt, Denner:2019vbn}) associated with the external
photon, and are added a posteriori to the virtual corrections.
The triangular fermionic loops form a self-contained class of UV and IR
finite diagrams and have not been included, as they cancel out for massless
fermions once all fermions of the same generation are considered, and the
mass contributions are known and negligible for our
study~\cite{Hagiwara:1991xy}.

It is, however, important to note that the use of numerical libraries for the
calculation of the virtual corrections to $pp \to Z\ph$ can present numerical
stability issues when considering regions of the $\pt$ spectrum where the $Z$
transverse momentum is very small compared to the hard process scale (in this
case, of order $\MZ$). For this reason, we have implemented an independent
calculation of the virtual matrix elements, that allowed us to derive
analytic formulae for the virtual correction in the singular regions, both in
the soft and the collinear regime, so increasing the precision and numerical
stability in the small $\pt$ limit.
}

We have computed analytically the virtual photonic corrections for the $pp
\to Z\ph$ process. The rationale behind this is that we want a code that is
perfectly stable also in the singular regions. Indeed, we have also computed
analytically the soft and collinear limits of the virtual amplitudes, as
Taylor expansion of the full analytic result, greatly increasing the
precision and numerical stability in the small $\pt$ limit, if compared to
any numerical calculation of the virtual amplitudes.

To check the calculation we have also computed the virtual corrections with
RECOLA, using its interface implemented in the {\tt POWHEG BOX RES} code, and
found very good agreement in kinematic configurations where the transverse
momentum of the $Z$ boson is sufficiently large to be away from any singular
region.

In the class of the virtual corrections, we have to consider also the
fermionic loops that enter in the counterterm of $\alpha(\mu)$ and in the
$\delta Z_\gamma$ factor (in the notation of Refs.~\cite{Denner:1991kt,
  Denner:2019vbn}) associated with the external photon, and are added a
posteriori to the virtual corrections, being proportional to the Born
amplitude.

To verify that the QED MiNLO code reaches the NLO accuracy for the fully
inclusive process $pp \to Z$, we compared the MiNLO results both to a
dedicated code, implemented in the {\tt POWHEG BOX RES} framework, and to a
modified version of the public {\tt Z\_ew-BMNNPV} code.\footnote{The modified
version of the code now includes processes where the $Z$ decays into
neutrinos.  In addition, also the $\gamma$-induced processes were added. The
implementation details and phenomenological impact of these diagrams on the
Drell-Yan precision physics will be documented in a dedicated publication.}

The electroweak parameters are computed in a hybrid scheme where the input
parameters are $\alpha(\mu)$ in the \MSB\ scheme, $\MW$ and $\MZ$.

\section{Validation and results}
\label{sec:res}
The goal of this section is to study how well the QED MiNLO formalism, based
on a calculation that has NLO QED accuracy for the process $pp\to Z\ph$,
approaches the NLO accuracy for the process $pp\to Z$, when fully inclusive
quantities are computed, and to validate some of the approximations that we
have discussed in Sec.~\ref{sec:minlo_QED}.

For this reason, we present a few numerical results obtained with the main
purpose of verifying the internal consistency of the MiNLO method we have
implemented and to quantify the degree of approximation it has.
It is important to emphasize from the outset that, due to the choice made for
the value of $\alpha$, in Sec.~\ref{sec:alpha}, and of the PDFs set that we
have used, in Sec.~\ref{sec:pdfs}, the absolute predictions for the cross
sections and distributions, as well as the relative NLO versus LO
corrections, are not realistic: these choices were made just to amplify the
effect of the QED corrections, so that to highlight any possible problem in
the formalism.
The conclusions we draw, regarding the agreement between the MiNLO results and the
corresponding ones for the totally inclusive $Z$ production, remain valid for
the actual value of the physical parameters.

\dontshow{
However, the information regarding the relationship between the
results obtained via MiNLO and the inclusive target, and those related to the
internal consistency of the calculation, particularly concerning the role of
the scales, remains valid.  In the following results, a generation cut is
applied on the neutrino-pair invariant mass at $40$~GeV.  While this cut
would be mandatory in the presence of final-state charged leptons due to
$s$-channel photon exchange, for neutrinos it serves solely to stabilize the
simulation by truncating the suppressed tail of the $Z$ boson Breit-Wigner
distribution. We have verified that removing this cut does not affect the
conclusions drawn in the following.
}

\begin{figure}[htb!]
  \begin{center}
    \includegraphics[width=0.49\textwidth]{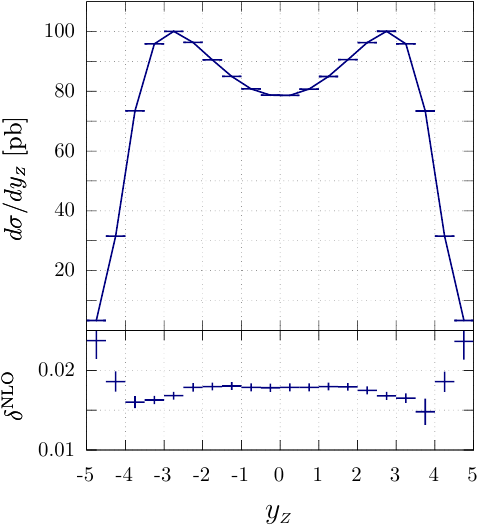}\ \
    \includegraphics[width=0.49\textwidth]{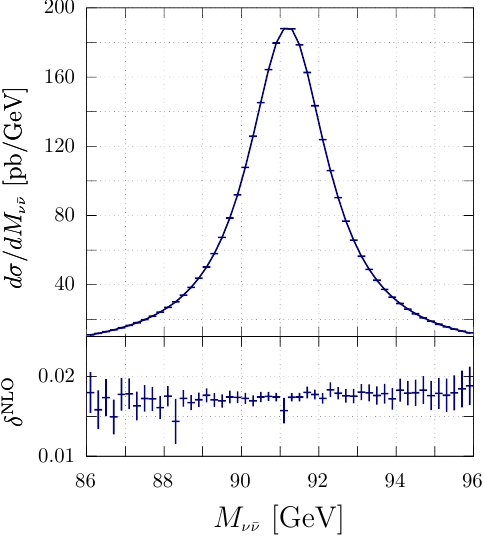}
  \end{center}
  \caption{Differential distributions for the inclusive process $pp\to Z$ as
    a function of the neutrino pair rapidity~(left) and invariant
    mass~(right) at NLO.  The lower panels show the ratio of
    the NLO over the LO corrections, with $\muf=\MZ$. The value of $\alpha$
    entering the corrections corresponds to $\alpha(\mur=\MZ)$, evolved from
    $\alpha(0)=\bar{\alpha}_0=0.04$. The PDF choice is detailed in
    Sect.~\ref{sec:pdfs}. The $\gamma$-induced contributions enter the
    calculation at NLO.}
  \label{fig:lo}
\end{figure}
The first thing to asses is the size of the NLO corrections with respect to
the LO, for the fully inclusive process, i.e.~$Z$ boson production in
Drell-Yan, with the $Z$ decaying into neutrinos.
In Fig.~\ref{fig:lo} we plot the rapidity and the invariant mass of the
neutrino pair, and $\delta^{\rm\sss NLO}$ is the ratio between the NLO over
the LO differential cross section.  The plots show that the effects are
positive and of the order of 2\%. The corrections are flat across the ranges
used in the figures, although, at large absolute rapidities, the corrections
tend to increase.  It should be noted that the power of $\alpha$ connected
with the photonic radiation for the inclusive Drell-Yan is evaluated in the
\MSB{} scheme, at the scale $\mur=\MZ$, as described in Sec.~\ref{sec:alpha}.

\begin{table}[ht]
\centering
\begin{tabular}{|l|c|c|c|c|c|c|}
%  \hline
\toprule
$\MZ/\ptcut$ &
$\sigma^{\sss >}_{\mathcal{O}(\alpha)}$ & $\sigma^{\sss <}_{\mathcal{O}(\alpha)}$ &
$\sigma^{\sss >}_{\mathcal{O}(\alpha)}+\sigma^{\sss <}_{\mathcal{O}(\alpha)}$  &
$\sigma_{\sss \rm  NLO}^{\sss \rm DY}$ &
$\sigma_{\sss \rm  LO}^{\sss \rm DY}$ &
 $\delta_{\alpha}$ \\ 
\midrule
  1 & 2.0191(5) & 730.4(1) & 732.5(1) & 735.346(7) & 722.818(4) & -0.0039(2) \\ 
  5 & 20.353(5) & 714.7(1) & 735.0(1) & 734.859(7) & 708.472(4) & 0.0003(2) \\
 10 & 36.35(1) & 698.3(1) & 734.7(1) & 734.437(5) & 702.774(4) & 0.0003(2) \\
 20 & 57.57(2) & 676.4(1) & 734.0(1) & 733.63(1) & 697.359(4) & 0.0005(3) \\
 50 & 93.456(3) & 639.3(1) & 732.7(1) & 732.476(7) & 690.638(4) & 0.0004(2) \\
100 & 126.181(4) & 605.1(1) & 731.3(1) & 731.24(1) & 685.897(4) & 0.0001(2) \\
200 & 163.566(5) & 565.9(1) & 729.5(1) & 729.395(5) & 681.490(4) & 0.0001(2) \\
\bottomrule
\end{tabular}
\caption{
  Integrated cross sections, in pb, obtained with the $\mathcal{O}(\alpha)$
  expansion of the MiNLO formulae for the regions below ($\sigma^{\sss
    <}_{\mathcal{O}(\alpha)}$) and above
  ($\sigma^{\sss>}_{\mathcal{O}(\alpha)}$) $\ptcut$, for some values of the
  technical cut ranging from $\MZ$ to $\MZ/200$.  The factorization scale
  $\muf$ in both the MiNLO and the target predictions is always set to
  $\ptcut$, while the power of the radiation $\alpha$ is set to
  $\alpha(\mu=\MZ)$.  The sum of the cross sections above and below $\ptcut$
  is compared to the cross section for the target process $pp\to Z$ at NLO.
  The last column shows the relative difference between the MiNLO prediction
  expanded at order $\mathcal{O}(\alpha)$ and $\sigma_{\sss \rm NLO}^{\sss
    \rm DY}$, as defined in eq.~(\ref{eq:delta_DYNLO}). }
\label{tab:expanded}
\end{table}
In order to estimate the effect of having neglected the finite contribution
$R_f$ in the region below $\ptcut$ in eq.~(\ref{eq:sigma<}), we have compared
the expansion of the MiNLO formulae at first order in $\alpha$, in both the
regions, below~(see eq.~(\ref{eq:integrated_sigma<})) and above~(see
eq.~(\ref{eq:sigma>})) the technical cut $\ptcut$, and we have compared their
sum to the fixed-order prediction for Drell-Yan at NLO accuracy.  The
differences between these two contributions gives the size of the neglected
term $R_f$.  In Table~\ref{tab:expanded}, we then present the total cross
sections obtained with the expanded MiNLO formulae, both in the regions
below, $\sigma^{\sss <}_{\mathcal{O}(\alpha)}$, and above, $\sigma^{\sss
  >}_{\mathcal{O}(\alpha)}$, the technical cut $\ptcut$.  Their sum is
compared with the prediction for Drell-Yan at NLO accuracy (the results
obtained for Drell-Yan at LO are also reported for completeness).  In the
last column we report the value of
\begin{equation}
\label{eq:delta_DYNLO}
%\delta^{\rm \sss DYNLO}
\delta_\alpha
\equiv \frac{\sigma^{\sss >}_{\mathcal{O}(\alpha)}+\sigma^{\sss
  <}_{\mathcal{O}(\alpha)}}{\sigma_{\sss \rm  NLO}^{\sss \rm DY}}  -1\,.
\end{equation}
According to the discussion in Sec.~\ref{sec:alpha}, the radiation $\alpha$
in all the cross sections in Table~\ref{tab:expanded} is computed in the
\MSB{} scheme at $\mur=\MZ$. We note that any variations of this scale in the
predictions obtained with MiNLO in the above and below $\ptcut$ regions
would, in any case, be of order $\mathcal{O}(\alpha^2)$.
The cross sections for the $\pt$ spectrum region above $\ptcut$ is the LO
contribution to $pp \to Z \ph$, while, for the region below $\ptcut$,
according to eq.~(\ref{eq:integrated_sigma<}), the $\mathcal{L}^{(0)}$ and
$\mathcal{L}^{(1)}$ expansion of the luminosity in eqs.~(\ref{eq:ll0})
and~(\ref{eq:ll1}) appear, along with the $\mathcal{O}(\alpha)$ expansion of
the Sudakov form factor.  In the calculation of these cross sections, the
PDFs are evaluated at $\muf=\ptcut$.

From Table~\ref{tab:expanded}, it can be observed that, for all considered
values of $\ptcut$ except $\MZ$, the MiNLO formula expanded to
$\mathcal{O}(\alpha)$ is in agreement with the inclusive NLO target, within
the numerical error.  This allows us to conclude that the
$\mathcal{O}(\alpha)$ regular term $R_f$, in the region below $\ptcut$ is
negligible, and it is reasonable to assume that the $\mathcal{O}(\alpha^2)$
terms are negligible too. This supports the procedure discussed in
Sec.~\ref{sec:minlo_QED}.
The table also shows the dependence of the total cross sections on the
factorization scale $\muf$: the effect of varying $\muf$, while being
considerably reduced in the transition from LO to NLO, becomes progressively
more pronounced as $\ptcut$ decreases. This is a consequence of the PDFs we
have adopted, whose $\muf$ dependence is rather steep in the region below the~GeV.

\begin{table}[h!t]
\centering
\begin{tabular}{|l|c|c|c|c|c|c|}
\toprule
$\MZ/\ptcut$ & $\sigma^{\sss >}$& $\sigma^{\sss >}_{\rm \sss LO}$ & $\sigma^{\sss
  <}$ & $\sigma^{\sss >}+\sigma^{\sss <}$ &  $\delta^{\rm \sss FO}_{\rm \sss MiNLO}$
&  $\delta^{\rm \sss FO}_{\rm \sss MiLO}$ \\
\midrule
  1 & 2.249(2)   & 2.018(2)  & 730.18(4) & 732.43(4) & -0.0039(1) & -0.00427(5) \\
  5 & 21.80(1)   & 20.60(1)  & 714.39(4) & 736.19(4) &  0.0011(1) & -0.00047(5) \\
 10 & 38.27(2)   & 36.48(1)  & 698.46(4) & 736.73(4) &  0.0019(1) & -0.00054(5) \\
 20 & 59.30(3)   & 56.96(2)  & 677.82(3) & 737.13(5) &  0.0024(1) & -0.00077(5) \\
 50 & 93.13(4)   & 90.02(3)  & 644.52(3) & 737.66(5) &  0.0031(1) & -0.00102(6) \\
100 & 122.29(6)  & 118.60(3) & 615.77(3) & 738.07(6) &  0.0037(1) & -0.00131(6) \\
200 & 153.58(8)  & 149.33(4) & 584.86(3) & 738.45(8) &  0.0042(2) & -0.00156(7) \\
\bottomrule
\end{tabular}
\caption{MiNLO predictions for the integrated cross sections, in pb, below
  ($\sigma^{\sss <}$) and above ($\sigma^{\sss >}$ and $\sigma^{\sss >}_{\rm
    LO}$) the separator $\ptcut$, for some values of the technical cut
  ranging from $\MZ$ to $\MZ/200$.  The cross section $\sigma^{\sss >}_{\rm
    LO}$ represents the integrated cross section above $\ptcut$, computed
  with the MiNLO and MiLO methods~(see the text for more details).  The last
  two columns show the relative difference between $\sigma^{\sss <} +
  \sigma^{\sss >}$ ($\sigma^{\sss <} + \sigma^{\sss >}_{\rm LO}$) and the
  (inclusive) total cross section for the reference process $pp\to Z$ at NLO.
  In this case, the target process is computed for $\muf=\MZ$ and its total
  cross section is given in eq.~(\ref{eq:sig_NLO}).}
\label{tab:minlo}
\end{table}
In Table~\ref{tab:minlo} we collect the value of the cross sections computed
with the MiNLO method, in the region below and above $\ptcut$, their sum, and
the relative difference between the MiNLO predictions and the reference
target.  Note that the target now is the NLO Drell-Yan cross
section, computed with $\muf=\MZ$, and with the electromagnetic coupling
$\alpha(\mur)$ computed in the \MSB{} scheme with $\mur=\MZ$.
The reason for this choice of scales is that the MiNLO formulae effectively
perform the evolution of the contributions evaluated at the scale
$\muf=\mur=\pt$ up to the hard scale of the inclusive reference process (in
this case, $\MZ$).
The target cross section is $\sigma_{\sss \rm  NLO}^{\sss \rm DY}$ in
Table~\ref{tab:expanded}, in the first line, i.e.
\begin{equation}
\label{eq:sig_NLO}
\sigma_{\sss\rm ref} = 735.346\pm 0.007 {\rm \ pb}\,.
\end{equation}
The relative difference between the MiNLO predictions and the reference
target is then defined as
\begin{equation}
\label{eq:delta_minlo}
\delta^{\rm \sss FO}_{\rm \sss MiNLO} = \frac{\sigma^{\sss >}+\sigma^{\sss
    <}}{\sigma_{\sss\rm ref}} -1\,.
\end{equation}
Table~\ref{tab:minlo} also shows a second set of results, where the cross
section above $\ptcut$ is computed in a LO variant of the MiNLO method, that
we dub MiLO.  Referring to eq.~(\ref{eq:STDminlo}), in the MiLO version, only
the Born contribution to the $pp \to Z\ph$ process contributes in the curly
brackets, and the Sudakov form factor in front is evaluated including only
the $A_1$ and $B_1$ terms in eq.~(\ref{eq:AB}).

In order to better interpret the numerical results, it might be useful to
recall the expression for the singular part of the cross section
$\mathd\sigmas$ in eq.~(\ref{eq:sigma_sing}) in the MiNLO
method~\cite{Hamilton:2012rf}, written as in eq.~(\ref{eq:ptres1}),
\begin{equation}
  \label{eq:AK}
  \frac{\mathd\sigmas}{\mathd\PhiB \mathd\pt} =
  e^{-\tilde{S}\(p_{\scaleto{\rm T}{3pt}}\)} \, \, \frac{\mathd
    |M^{\scriptscriptstyle\rm Z}|^2_{ij} }{\mathd\PhiB} \, \,
  \frac{2}{\pt} \sum_{n=1}^2 \,\, \sum_{m=0}^1 \( \frac{\alpha(\pt)}{2\pi} \)^n
   {}_nE_m \log^m \! \(\frac{Q^2}{\pt^2}\)\,,
\end{equation}
where the coefficients $ _n E_m$, using the same notation as in
eqs.~(\ref{eq:ll0}) and~(\ref{eq:ll1}), read
\begin{eqnarray}
  \label{eq:E_coeffs}
_1E_1 & = &\Aone  f_i^{[a]} f_j^{[b]},
\nonumber
\\ 
_1 E_0 & = &\Bone f_i^{[a]} f_j^{[b]} + \( P^{(0)}_{ik} \otimes f_k \)\!\rule{0pt}{1.9ex}^{[a]}  f_j^{[b]} + f_i^{[a]}
\(P^{(0)}_{jk} \otimes f_k \)\!\rule{0pt}{1.9ex}^{[b]},
 \nonumber \\
_2 E_1 & = & \frac{1}{2}\Atwo  f_i^{[a]} f_j^{[b]} +\Aone \lq  \frac{1}{2}H f_i^{[a]} f_j^{[b]} +
\(C_{ik} \otimes f_k \)^{[a]}  f_j^{[b]}  \rq + \{i \leftrightarrow j\} ,
\nonumber \\ 
_2 E_0 & = & \frac{1}{2} \Btwot f_i^{[a]} f_j^{[b]} +\( \Bone -2
\pi \beta_0^{\rm \sss QED}\) \lq
\frac{1}{2}H f_i^{[a]} f_j^{[b]} + \(C_{ik} \otimes f_k \)\!\rule{0pt}{1.9ex}^{[a]}  f_j^{[b]}  \rq 
\nonumber \\ 
&&\mbox{} + H \(P^{(0)}_{ik} \otimes f_k \)\!\rule{0pt}{1.9ex}^{[a]}  f_j^{[b]}
    + \( C_{ik} \otimes P^{(0)}_{kl}  \otimes f_l\)\!\rule{0pt}{1.9ex}^{[a]} f_j^{[b]} + \(C_{ik} \otimes f_k \)\!\rule{0pt}{1.9ex}^{[a]}
    \(P^{(0)}_{jl} \otimes f_l \)\!\rule{0pt}{1.9ex}^{[b]} 
    \nonumber \\
&&\mbox{} + \( P^{(1)}_{ik} \otimes f_k\)\!\rule{0pt}{1.9ex}^{[a]} f_j^{[b]}
+ \{i \leftrightarrow j\}\, .  
\end{eqnarray}
The terms $_1E_1$ and $_1 E_0$ represent the Born-level contribution to the
process $pp\to Z\ph$ and correspond to the
$\mathcal{O}(\alpha)$ contribution of $\mathd \{ \exp[-\tilde{S}(\pt)] \lum\}
/ \mathd \pt$ in eq.~(\ref{eq:sigma_sing}).
So, as long as we restrict ourselves to the singular part of the cross
section, MiLO corresponds to eq.~(\ref{eq:AK}), where the Sudakov form factor
contains only $\mathcal{O}(\alpha)$ terms and where only the terms $_1 E_1$
and $_1 E_0$ are non vanishing.  The terms $_2 E_1$ and $_2 E_0$ are the
singular limit of the real and virtual corrections to $pp\to Z\ph$, from
which the first-order expansion
of the exponential in eq.~(\ref{eq:AK}) has been subtracted.  This
subtraction is necessary to eliminate the $\log^3(Q^2/\pt^2)$ and
$\log^2(Q^2/\pt^2)$ terms, otherwise present in the expression of the $Z$
boson $\pt$ spectrum at NLO~\cite{Arnold:1990yk}.

To interpret the results for $\delta^{\rm \sss FO}_{\rm \sss MiNLO}$ in
Table~\ref{tab:minlo}, we need to consider also the following issues:
\begin{enumerate}
\item
The cancellation of the $\log^m(Q^2/\pt^2)$ ($m=2,3$) terms is purely
numerical in MiNLO and can become critical when studying relative differences
at the per mille level (as in this case).

\item
The terms containing the one-loop corrections to the Altarelli-Parisi
kernels, $P^{(1)}_{ij}$ in eq.~(\ref{eq:E_coeffs}), arise from the collinear
limit of the double-radiation contributions and should correspond to the
$\ord{\alpha^2}$ component of the PDF evolution equation.  The HOPPET code we
have used does not provide QED evolution of the PDFs at this order.  So the
PDF evolution does not include $\mathcal{O}(\alpha)$ corrections to the
Altarelli-Parisi kernels, and the expression in eq.~(\ref{eq:AK}) is no
longer exactly equal to the derivative of $\exp[-\tilde{S}(\pt)] \lum$ with
respect to $\pt$ of eq.~(\ref{eq:sigma_sing}).
This gives rise to a mismatch in the sum of the cross sections $\sigma^{\sss
  >}$ and $\sigma^{\sss <}$ that depends on the $\ptcut$ cutoff.

\item
In addition to terms singular in the $\pt \to 0$ limit, the MiNLO cross
section contains also regular contributions of order $\alpha$ and $\alpha^2$
(the $R_f$ term).  The contribution at order $\alpha$ is necessary to
reproduce the NLO corrections to the totally inclusive Drell-Yan process
$pp\to Z$.  The contribution at order $\alpha^2$ is necessary to reproduce
the NLO corrections for $pp\to Z\ph$. But when comparing MiNLO and the target
cross section, this contribution is effectively a spurious higher-order
effect, that can contribute to give some discrepancies between MiNLO and the
target.

\end{enumerate}
\dontshow{
 for $\delta^{\rm \sss FO}_{\rm \sss
  MiLO}$, defined as $\delta^{\rm \sss FO}_{\rm \sss MiNLO}$ in
eq.~(\ref{eq:delta_minlo}) but with $\sigma^{\sss >}$ replaced by
$\sigma^{\sss >}_{\rm \sss LO}$, }
On the other hand, the MiLO results are not affected by the aforementioned
issues.
However, they contain neither the $\mathcal{O}(\alpha^2)$ term associated
with the derivative of the Sudakov form factor~(see the first term in
eq.~(\ref{eq:D2})), nor the terms related to $\alpha(\pt)\,[\lum]^{(1)}$ and
its derivative with respect to $\pt$, which are nevertheless necessary to
achieve NLO accuracy for observables inclusive in the $Z$ boson transverse
momentum. The former contribution consists in the $\Atwo$ and $\Btwot$ terms
of the $_2 E_1$ and $_2 E_0$ coefficients in eq.~(\ref{eq:E_coeffs}) and it
can be written as the difference
\begin{equation}
  \label{eq:miloapprox1}
  \Delta = 
\frac{\mathd}{\mathd \pt} \lg\exp\lq -\tilde{S}(\pt)\rq  \lq\mathcal{L}(\pt)\rq^{(0)}\rg 
-
\frac{\mathd}{ \mathd \pt} \lg\exp\lq -\tilde{S}_1(\pt)\rq \lq
\mathcal{L}(\pt)\rq^{(0)}\rg ,
\end{equation}
where $\tilde{S}_1$ is a short-hand notation for $\tilde{S}$, computed
including only the $\Aone$ and $\Bone$ coefficients. We can further
manipulate $\Delta$ to get
\begin{eqnarray}
  \Delta 
&=& - \exp\lq -\tilde{S}(\pt) \rq \lg \lq \frac{\mathd\tilde{S}(\pt)}{\mathd
  \pt} -\frac{\mathd\tilde{S}_1(\pt)}{\mathd \pt}
\exp\!\( \tilde{S}(\pt)  -\tilde{S}_1(\pt)\)\rq 
\lq\mathcal{L}(\pt)\rq^{(0)} \right.
\nonumber\\
&& \phantom{-\exp\lq -\tilde{S}(\pt)\rq \big\{}+ \frac{\mathd}{\mathd
  \pt}\(\lq\mathcal{L}(\pt)\rq^{(0)}\)  
\lq 1- \exp\!\( \tilde{S}(\pt)  -\tilde{S}_1(\pt)\) \rq ,
\end{eqnarray}
and, since $\exp\( \tilde{S}(\pt) -\tilde{S}_1(\pt)\) = 1 + \ord{\alpha^2}$
and $\mathd (\lq\mathcal{L}(\pt)\rq^{(0)})/ \mathd \pt = \ord{\alpha}$,
using eq.~(\ref{eq:DSud}) for the derivative of the Sudakov form factor, we
get
 \begin{equation}
 \Delta = \exp\!\lq -\tilde{S}(\pt)\rq \frac{2}{\pt}
 \(\frac{\alpha(\pt)}{2\pi}\)^2
 \lg A^{(2)}\log\frac{Q^2}{\pt^2} + \tilde{B}^{(2)}\rg \lq\mathcal{L}(\pt)\rq^{(0)} + \ord{\alpha^3}.
\end{equation}
In a similar way, it can be shown that the remaining terms of $_2 E_1$ and
$_2 E_0$, omitting the $P^{(1)}_{ik}$ contribution that is missing in the
evolution of the PDF set, as discussed in Sec.~\ref{sec:pdfs}, can be cast in
the form\footnote{Here we also need to use
\begin{equation}
\frac{\mathd\alpha(\pt)}{\mathd \pt} = -\frac{2}{\pt} \beta_0^{\rm \sss QED}
\alpha^2(\pt) + \ord{\alpha^3}.
\end{equation}
}
\begin{equation}
  \label{eq:miloapprox2}
\frac{\mathd}{\mathd \pt} \lg\exp\lq-\tilde{S}_1(\pt)\rq
\frac{\alpha(\pt)}{2\pi} \lq \mathcal{L}(\pt)\rq^{(1)}\rg. 
\end{equation}
The missing contributions to the integrated cross-section in the MiLO
approximation can thus be estimated analogously to the procedure used for
$\sigma^{\sss <}$ in eq.~(\ref{eq:integrated_sigma<}), i.e.~by integrating
the exact differentials in eqs.~(\ref{eq:miloapprox1})
and~(\ref{eq:miloapprox2}), between the technical cutoff $\ptcut$ and the
hard process scale.

By including also these contributions in $\sigma^{\sss >}_{\rm \sss LO}$, the
relative discrepancy $\delta^{\rm \sss FO}_{\rm \sss MiLO}$ (defined as
$\delta^{\rm \sss FO}_{\rm \sss MiNLO}$ in eq.~(\ref{eq:delta_minlo}) but
with $\sigma^{\sss >}$ replaced by $\sigma^{\sss >}_{\rm \sss LO}$), shown in
the last column of Table~\ref{tab:minlo}, reduces to approximately $0.5$ per
mille for $\ptcut$ below 20~GeV, confirming the fact that the implementation
is correct and the formalism we have proposed is sound.

\begin{figure}[htb!]
  \begin{center}
    \includegraphics[width=0.49\textwidth]{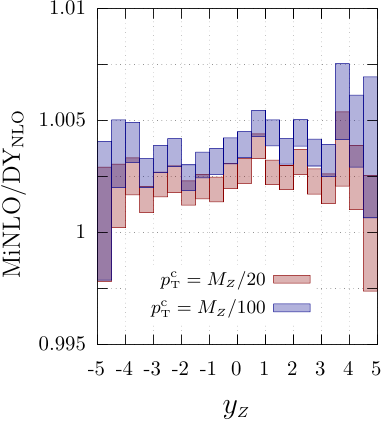}\ \
    \includegraphics[width=0.49\textwidth]{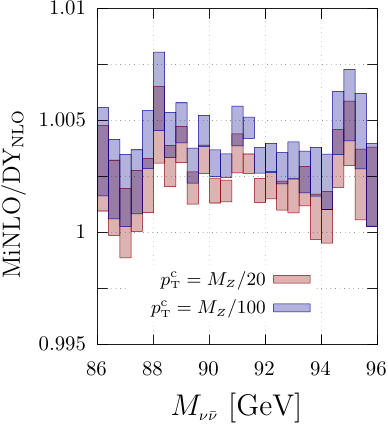}
  \end{center}
  \caption{Ratio between the MiNLO predictions for $pp\to Z\ph$ and the
    target process $pp\to Z$ at NLO QED as a function of the neutrino pair
    rapidity~(left) and invariant mass~(right). The predictions for the
    Drell-Yan process are the same ones shown in Figure~\ref{fig:lo}, while
    the MiNLO results correspond to the sum of the contributions above and
    below $\ptcut$, for the technical cut equal to $\MZ/20$ and $\MZ/100$.}
  \label{fig:ratios}
\end{figure}
Turning now to more differential quantities, in Fig.~\ref{fig:ratios} we show
the MiNLO differential cross section for the inclusive production of a $Z$
boson, as a function of the rapidity and of the invariant mass of the two
neutrinos, for two values of the technical cut $\ptcut$: $\MZ/20$ and
$\MZ/100$.  Similarly to the conclusions on the total cross sections derived
from Table~\ref{tab:minlo}, the agreement between the MiNLO differential
cross sections and the NLO target is of the order of a few per mille, to be
compared to the size of the NLO corrections with respect to the LO one,
depicted in Fig.~\ref{fig:lo}, of the order of 2\%,
%placing the approximation of the MiNLO method at 20\% of the correction,
for the augmented value of $\alpha$ that we are using in this section.

\begin{table}[ht]
\centering
\begin{tabular}{|l|c|c|c|c|}
\toprule
$\MZ/\ptcut$ & $\sigma^{\sss >}$ & $\sigma^{\sss <}$ & $\sigma^{\sss
  >}+\sigma^{\sss <}$ &  $\delta^{\rm \sss FO}_{\rm \sss MiNLO}$  \\
\midrule
  1 & 0.1172(1) &   651.41(1)  & 651.53(1)  & -0.00249(2)  \\
  5 & 1.5765(4) &   651.41(1)  & 652.99(1)  & -0.00026(2)  \\
 10 & 3.245(1)  &   649.88(1)  & 653.12(1)  & -0.00005(2)  \\
 20 & 5.739(1)  &   647.44(1)  & 653.18(1)  &  0.00004(2)  \\
 50 & 10.360(1) &   642.86(1)  & 653.21(1)  &  0.00009(2)  \\
100 & 14.845(2) &   638.389(4) & 653.235(5) &  0.00011(1)  \\
200 & 20.165(3) &   633.090(4) & 653.255(5) &  0.00015(1)  \\
\bottomrule
\end{tabular}
\caption{MiNLO predictions for the integrated cross sections, in pb, below
  ($\sigma^{\sss <}$) and above ($\sigma^{\sss >}$) the technical cut
  $\ptcut$, ranging from $\MZ$ to $\MZ/200$, and for realistic value of the
  electromagnetic coupling $\alpha = \bar{\alpha}_0=0.008$. In the last
  column, the relative difference as defined in eq.~(\ref{eq:delta_minlo}),
  with $\sigma_{\sss\rm ref} = 653.160(6)$~pb, the total cross section for
  the reference process $pp\to Z$ at NLO computed for $\muf=\MZ$.  In this
  table, the PDF set used has been consistently evolved with the same choice
  of the value for $\alpha$. Note that the LO prediction for the reference
  process is $650.73(1)$~pb, so that the relative NLO corrections for $pp\to
  Z$ amount to $3.74(2)$ per mille.  }
\label{tab:minloa0}
\end{table}

If we use instead the physical value of $\alpha$, the corrections are reduced
by about six times, and this also reduces the difference between the
predictions for the target process at NLO and the ones for $Z\ph$ production,
obtained with MiNLO, as well as the residual $\ptcut$ dependence of the
results.  This is shown in Table~\ref{tab:minloa0}, where the calculation of
Table~\ref{tab:minlo} has been repeated using an initial value for the
running of $\alpha$ equal to $\bar{\alpha}_0=0.008$. In this setup, the NLO
total cross section for the inclusive $Z$ production turns out to be
653.160(6)~pb, and the LO one $650.73(1)$~pb, so that the NLO corrections
reduce to approximately $3.74(2)$ per mille.

However, the absolute difference (and, to a good approximation, the relative
one as well) between the MiNLO predictions and the NLO-accurate Drell-Yan
ones scales as $\alpha^2$, resulting in a relative difference with respect to
the target of the order of 0.1\textperthousand, as shown in the last column
of Table~\ref{tab:minloa0}.  Since the PDFs in this case were evolved using a
consistently modified $\alpha$, one cannot expect the scaling of the effects
to be determined exactly by the ratio of the $\alpha$ values
used. Nevertheless, it is possible to observe that the numbers in the last
column of Table~\ref{tab:minloa0} scale approximately as the square of the
ratio of the $\alpha$ values, when compared to those in the last column of
Table~\ref{tab:minlo}.

With a view to jointly addressing combined QCD and QED corrections, the QCD
theory would constrain the choice of $\ptcut$ to values no smaller than $\sim
\!1$~GeV.  We can therefore take the results corresponding to $\ptcut
\sim\!1$~GeV (next-to-last row in Table~\ref{tab:minloa0}) as a reasonable
estimate of the accuracy of the MiNLO method variant. We can then conclude
this section by saying that, by applying the the MiNLO method as presented in
this paper, we get an accuracy within the $0.01\%$ between MiNLO and the NLO
target, for a realistic choice of the value of the QED coupling $\alpha$.

\section{Conclusions}
\label{sec:conclusions}
The electroweak physics program at the (HL-)LHC, particularly the precision
measurements based on template fits to Drell-Yan processes, demands Monte
Carlo tools that incorporate frontier calculations of both QCD and EW
radiative corrections to minimize theoretical systematic
uncertainties. Currently, state-of-the-art event generators for Drell-Yan
production achieve NNLO + parton shower~(PS) accuracy in QCD, while EW
effects are available in generators featuring NLO QCD + NLO EW accuracy
matched to QED and QCD parton showers for the inclusive process. On one hand,
a framework capable of simultaneously treating NNLO QCD and NLO EW
corrections matched to their respective parton showers is still missing. On
the other hand, no tool currently provides NLO EW+PS accuracy for both the
inclusive process and DY production in association with resolved radiation.
This latter aspect is becoming increasingly important as the growing LHC
statistics allow experimental collaborations to perform differential
measurements over radiation-sensitive variables, such as the vector-boson
transverse momentum.

The MiNNLO$_{\text{PS}}$ method, designed to achieve NNLO QCD accuracy for an
inclusive process, starting from the simulation of the corresponding
radiative process at NLO, represents a natural starting point for developing
a DY event generator with NNLO QCD + NLO EW + PS accuracy.

In this work, we take the first steps in this direction by presenting a
possible formulation of the MiNLO method for QED corrections. We follow the
original QCD formulation of MiNLO as closely as possible to highlight the
specific features that arise in the QED case. To this end, we adopted $pp \to
Z(\to \nu\bar{\nu})$ as the inclusive reference process, in order to have
only initial-state radiation effects.  We start from the QED corrections to
$pp \to Z\ph$ (with $\ph = \gamma, q, \bar{q}$), including both quark and
photon-initiated processes.
We have presented the abelianization of the MiNLO formulae and we have shown
that the modified couplings and coefficients in the Sudakov form factor have
a deep impact on the position of its peak, locating it at such a small value
of the transverse momentum of the vector boson, where any numerical
evaluation of the corresponding amplitudes turns out to be totally
unreliable.
To address this issue, we proposed a strategy for the computation of the
small transverse-momentum differential cross section, exploiting the
analytical properties of the MiNLO formula. This led to the introduction of a
technical cutoff, $\ptcut$, that we have used to separate the differential
cross section into two contributions, neglecting power-suppressed terms in
the small transverse-momentum region.

In addition, we have discussed the treatment of the coupling associated with
radiative emission and the repercussions of the running of electromagnetic
coupling $\alpha$ on the calculation of the radiative corrections.
We have addressed the role of the parton distribution functions and the
interplay between the QED evolution of the PDFs and the MiNLO formulae.  We
have presented the validation of the method, along with several numerical
results, in the setup featuring only the first-generation of quarks and an
enhanced value of the electromagnetic coupling, in order to emphasize
physical effects and potential discrepancies with respect to the expected
behavior of the MiNLO formulae.

We have also quantified the uncertainties on the inclusive predictions
arising from the proposed formalism and missing PDF evolution terms, when
using a realistic value for $\alpha$. We have studied the dependence of the
differential cross section on the technical cut $\ptcut$, and we have found
that, for values of $\ptcut$ of the order of 1~GeV (typical cutoff value in
the QCD MiNLO applications), the uncertainty is at the $0.01\%$ level, for
initial-state radiation effects.

The study presented here is a necessary first step towards incorporating full
EW effects into the MiNNLO$_{\text{PS}}$ framework. Future developments will
involve extending this approach to final-state radiation, building upon the
MiNNLO formulation for $t\bar{t}$ production, achieving a combined treatment
of QCD and QED corrections and, finally, exploring the strategies to
incorporate full electroweak effects beyond the pure QED approximation.

\acknowledgments
We would like to thank L.~Buonocore, A.~Denner, G.~Ferrera, J.~Lindert,
P.F.~Monni, P.~Nason, L.~Rottoli, P.~Torrielli and S.~Uccirati for useful
discussions, and C.~Davies for providing us with a copy of her PhD thesis.
We would also like to thank V.~Bertone and S.~Carrazza for exchanges about
the APFEL code.  F.B.~and M.C.~would like to thank the kind hospitality of
the GGI Institute in Florence, Italy, during part of the work.

This work has been partially supported by the Italian Ministry of University
and Research (MUR), with EU funds (NextGenerationEU), through the PRIN2022
grant agreement Nr. 20229KEFAM (CUP H53D23000980006, I53D23001000006).

%% We acknowledge financial support, super-computing resources and support from
%% ICSC – Centro Nazionale di Ricerca in High Performance Computing, Big Data
%% and Quantum Computing – and hosting entity, funded by European Union –
%% NextGenerationEU.

This work is also partially supported by ICSC - Centro Nazionale di
Ricerca in High Performance Computing, Big Data and Quantum Computing,
funded by European Union - NextGenerationEU.

\bibliography{paper}

\end{document}

%% file: texmacros.tex
\newcommand{\mathd}{\mathrm{d}}

\newcommand\sss{\mathchoice%
{\displaystyle}%
{\scriptstyle}%
{\scriptscriptstyle}%
{\scriptscriptstyle}%
}
\newcommand{\ptcut}{p_{\sss\rm T}^{\sss\rm c}}
\newcommand{\pt}{p_{\sss\rm T}}

\def\beq{\begin{equation}}
\def\beqn{\begin{eqnarray}}
\def\eeq{\end{equation}}
\def\eeqn{\end{eqnarray}}

\def\lq{\left[} 
\def\rq{\right]} 
\def\rg{\right\}} 
\def\lg{\left\{} 
\def\({\left(} 
\def\){\right)} 
 
\newcommand\nf{n_{\rm f}}

\newcommand\as{\alpha_{\sss\rm S}}

\newcommand\CF{C_{\sss\rm F}}
\newcommand\CA{C_{\sss\rm A}}
\newcommand\TF{T_{\sss\rm F}}

\newcommand\NC{N_{\rm c}}

\newcommand\muf{\mu_{\sss\rm F}}
\newcommand\mur{\mu_{\sss\rm R}}

\newcount\minutes 
\newcount\scratch 
\def\timestamp{% 
\scratch=\time 
\divide\scratch by 60 
\edef\hours{\the\scratch} 
\multiply\scratch by 60 
\minutes=\time 
\advance\minutes by -\scratch 
---$\,$\hours:\null 
\ifnum\minutes< 10 0\fi 
\the\minutes}

\def\MZ{M_{\sss Z}}
\def\MW{M_{\sss W}}

\newcommand{\MSB}{\ensuremath{\overline{\rm MS}}}

\def\beeq{\begin{eqnarray}} 
\def\eeeq{\end{eqnarray}} 
\def\to{\rightarrow}

\newcommand\HCCF{\lq \HCCF H^{F} C_1 C_2 \rq_{c\bar{c};\,ab}}

\newcommand\Lum{{\cal L}}
\newcommand\ord[1]{\mathcal{O}\!\(#1\)}
\newcommand\lum{\Lum(\pt)}

\newcommand\dontshow[1]{}

\newcommand\PhiB{\Phi_{\sss Z}}

\newcommand{\abar}{\frac{\as}{2\pi}}
\newcommand{\abarmu}[1]{\frac{\as(#1)}{2\pi}}

\newcommand{\sigmas}{\sigma^{\rm sing}}
\newcommand\ph{{\cal A}}

\newcommand\Aone{A^{(1)}}
\newcommand\Bone{B^{(1)}}
\newcommand\Atwo{A^{(2)}}

\newcommand\Btwot{\tilde{B}^{(2)}}